\documentclass{webofc}
\usepackage[varg]{txfonts}   
\usepackage{lineno}
\usepackage{hyperref}

\begin{document}
\title{ Elliptic and triangular flow of (multi-)strange hadrons and $\phi$ mesons in BES-II energies at STAR}
%
%

\author{\firstname{Li-Ke} \lastname{Liu}\inst{1}\fnsep\thanks{\email{likeliu@mails.ccnu.edu.cn}} (for the STAR Collaboration)
}

\institute{Key Laboratory of Quark \& Lepton Physics (MOE) and Institute of Particle Physics, 
Central China Normal University, Wuhan 430079, China }

\abstract{
\normalsize
In these proceedings, we present the measurements of 
elliptic ($v_2$) and triangular ($v_3$) flow of (multi-)strange hadrons and $\phi$ mesons in 19.6 and 14.6 GeV Au+Au collisions from the STAR. The number of constituent quark (NCQ) scaling of 
$v_2$ and $v_3$ holds well at $\sqrt{ s_{\mathrm{ NN }} } $ = 19.6 GeV, which indicates
the collective flow is built up in the partonic stage. 
At these energies, the anti-particles show better NCQ scaling than the particles for both $v_2$ and $v_3$,
which may be caused by the different contributions from the produced and transported quarks.
}
\maketitle
\section{Introduction}
\label{intro}

At sufficiently high temperature and/or high density, quantum chromodynamics (QCD) predicts a transition 
from hadronic matter to deconfined partonic matter~\cite{Gross:1980br,Luo:2020pef}.
Results from top-RHIC and LHC energies indicate 
a new form of matter with small viscosity and high temperature, created in high-energy heavy ion collisions
~\cite{Bazavov:2011nk, Fukushima:2013rx}. 
Lattice QCD calculations predict that, the phase transition from hadronic matter to the QGP phase is a smooth crossover at vanishing baryon chemical potential ($\mu_{B}$) region. 
A first-order phase transition is expected at a finite baryon chemical potential region. 
Locating the first-order phase boundary with a critical point is essential in establishing the QCD phase diagram. 
It motivates the Beam Energy Scan (BES) program at RHIC, which covers energy $\sqrt{ s_{\mathrm{ NN }} } $ = 3.0 - 62.4 GeV corresponding baryon chemical potential 750 - 73 MeV. 
It can help us explore the QCD phase structure in the high baryon density region.

Anisotropies in particle momentum distributions relative to the reaction plane are referred to as anisotropic collective flow.
Elliptic flow coefficient, $v_2$, is sensitive to the dynamics at the early stages 
of the system evolution in heavy-ion collisions and equation of state of the medium~\cite{Voloshin:2008dg}. 
Triangular flow $v_3$ is particularly sensitive to the initial geometry fluctuations~\cite{Alver:2010gr}. 
The hadronic interaction cross sections of multi-strange hadrons 
and $\phi$ mesons are expected to be small~\cite{Mohanty:2009tz, Nasim:2013fb, Shi:2016elm}. 
Hence, the anisotropic flow of these hadrons provides 
information on the early stages of the high-energy collisions.
A hint of smaller $v_2$ for $\phi$ mesons compared to charged particles is observed for BES-I Au+Au collisions at $\sqrt{ s_{\mathrm{ NN }} } $ = 7.7 and 11.5 GeV~\cite{STAR:2013ayu}.

In these proceedings, with the enhanced statistics datasets, we report the measurements of $v_2$ and $v_3$ 
of (multi-)strange hadrons ($K^\pm$, $K_S^0$, $\Lambda$, $\bar{\Lambda}$, $\Xi$, $\bar{\Xi}^{+}$, $\Omega$, $\bar{\Omega}^{+}$) and $\phi$ mesons from BES-II Au+Au collisions 
at $\sqrt{ s_{\mathrm{ NN }} } $ = 14.6 and 19.6 GeV.

\section{Data sets and Analysis strategy}
\label{analysis}

Data samples for Au+Au collisions at $\sqrt{ s_{\mathrm{ NN }} } $ = 14.6 and 19.6 GeV 
taken in 2019 are used in the analysis.
The number of good events analyzed are 270$\times 10^6$ and 440$\times 10^6$ for 14.6 and 19.6 GeV, respectively.
The primary vertex position of each event along the beam direction, $V_{z}$, is selected to be within $\pm$ 70 cm 
from the center of the Time Projection Chamber (TPC).
To eliminate possible beam interactions with the vacuum pipe, the vertex along the radial direction, $V_r$, is selected to be smaller than 2 cm.
To select good quality tracks, $p_T$ > 0.2 GeV/$c$, and a distance of
closest approach (DCA) from vertex, DCA $\le$ 3 cm, and at least 15 space
points in the TPC acceptance are required. 
The particle identification at low transverse momentum is performed via 
their specific energy loss measured by the TPC. 
At intermediate and high transverse momenta, the particle identification is performed 
using the Time of Flight (TOF) detector. 
The strange hadrons such as $K_S^0$, $\Lambda$, $\bar{\Lambda}$, $\Xi^{-}$, ${\Xi}^{+}$, $\Omega^{-}$, and ${\Omega}^{+}$
are reconstructed using KF-Particle package~\cite{Banerjee:2020iab}.
The $\phi$ mesons are reconstructed via $K^+ K^-$ channel. 
The systematic uncertainties on the measurements are obtained by 
varying the above analysis cuts. The dominant sources of systematic uncertainties are the particle identification cuts and quality track selection cuts.


\section{Results and Discussions}
\label{results}

\begin{figure}[htbp]
    \centering
    \includegraphics[width=4in,keepaspectratio]{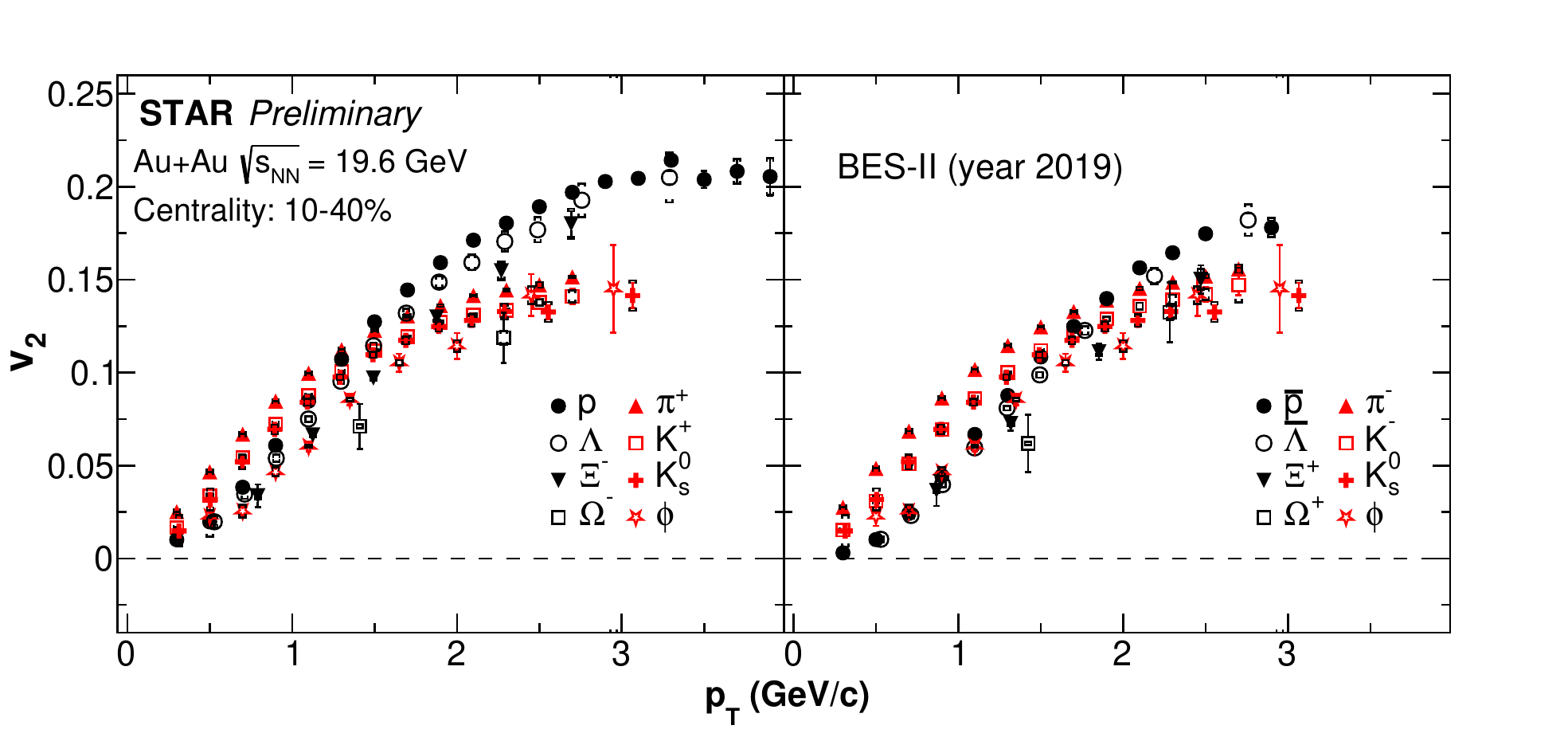}
    \caption{$p_T$ differential elliptic flow of the identified particles 
    in 10-40\% Au+Au collisions at $\sqrt{ s_{\mathrm{ NN }} }$ = 19.6 GeV. }
    \label{Fig:PIDv21040}
\end{figure}

Figure \ref{Fig:PIDv21040} shows the $p_T$ differential elliptic flow 
of the identified particles for mid-central (10-40\%) Au+Au collisions at $\sqrt{ s_{\mathrm{ NN }} }$ = 19.6 GeV.
The black markers show baryon $v_2$ results, and the red markers show meson $v_2$ results.
There is a clear mass ordering when $p_{T} < $ 1.5 GeV/$c$. 
It could be due to the interplay of radial expansion and anisotropic flow, 
which is consistent with hydrodynamics predictions~\cite{Borghini:2005kd}.
When $p_T > $ 1.5 GeV/$c$, the $v_2$ of particles is grouped according to the hadron type, 
called baryon-meson splitting, which indicates that the coalescence could be 
dominant process of hadronization in this $p_{T}$-region.

\begin{figure}[htbp]
    \centering
    \includegraphics[width=3.8in,keepaspectratio]{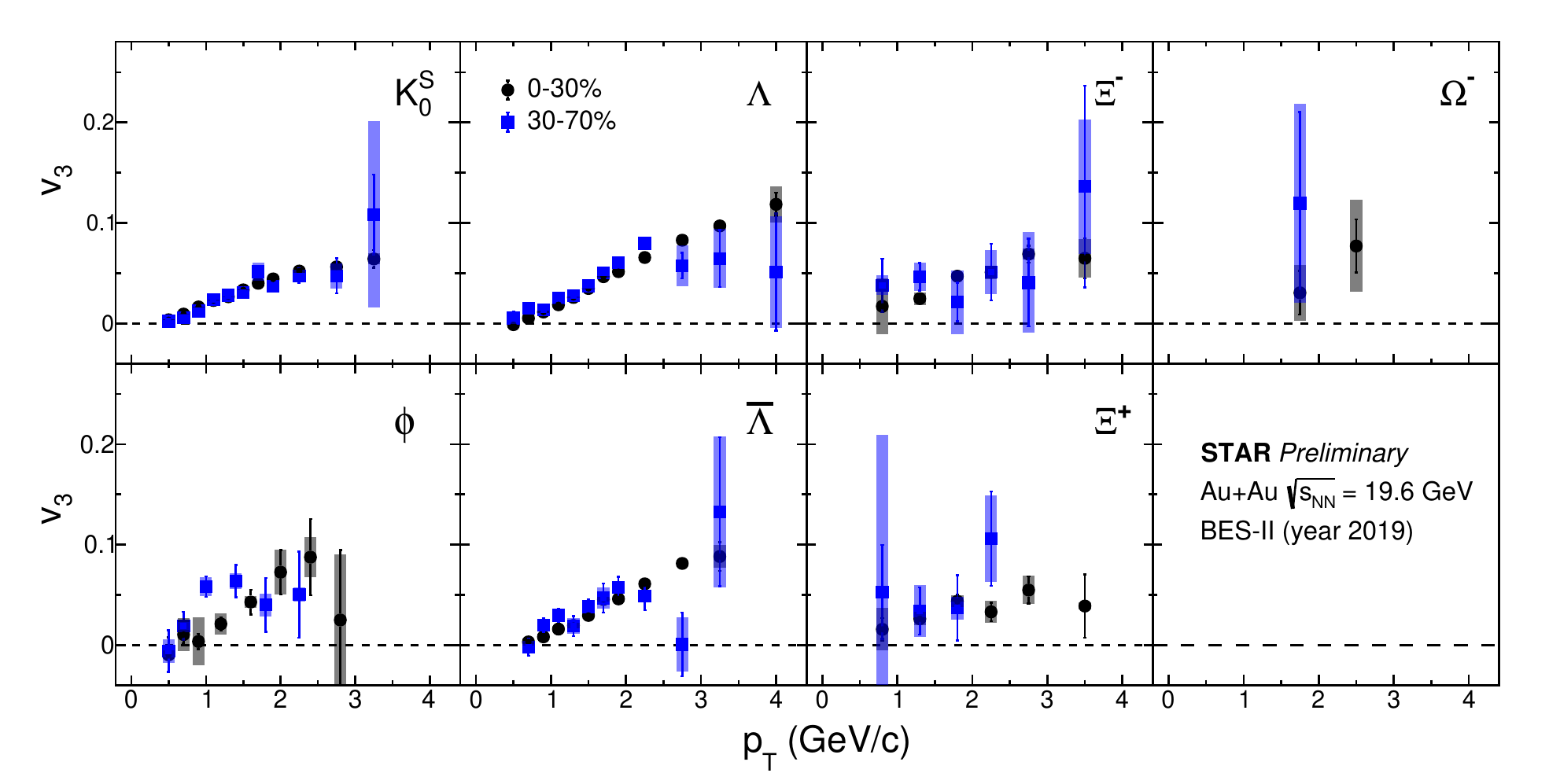}
    \caption{$p_T$ differential triangular flow of the identified particles 
    in 0-30\% and 30-70\% Au+Au collisions at $\sqrt{ s_{\mathrm{ NN }} }$ = 19.6 GeV. 
    The vertical bars and shaded bands are the statistical and systematic uncertainties, respectively.}
    \label{Fig:PIDv3}
\end{figure}

The $p_T$ differential triangular flow of the identified particles 
in 0-30\% and 30-70\% Au+Au collisions at $\sqrt{ s_{\mathrm{ NN }} }$ = 19.6 GeV
are shown in Fig. \ref{Fig:PIDv3}. 
The triangular anisotropy ($\epsilon_3$) in the initial collision geometry originates from event-by-event fluctuations. The data of $v_3$ show a weak centrality dependence, same as $\epsilon_3$~\cite{Alver:2010gr, Xiao:2011ti}. It indicates the event-by-event fluctuations are the dominant source for $v_3$.

\begin{figure}[htbp]
  \centering
  \includegraphics[width=3.8in,keepaspectratio]{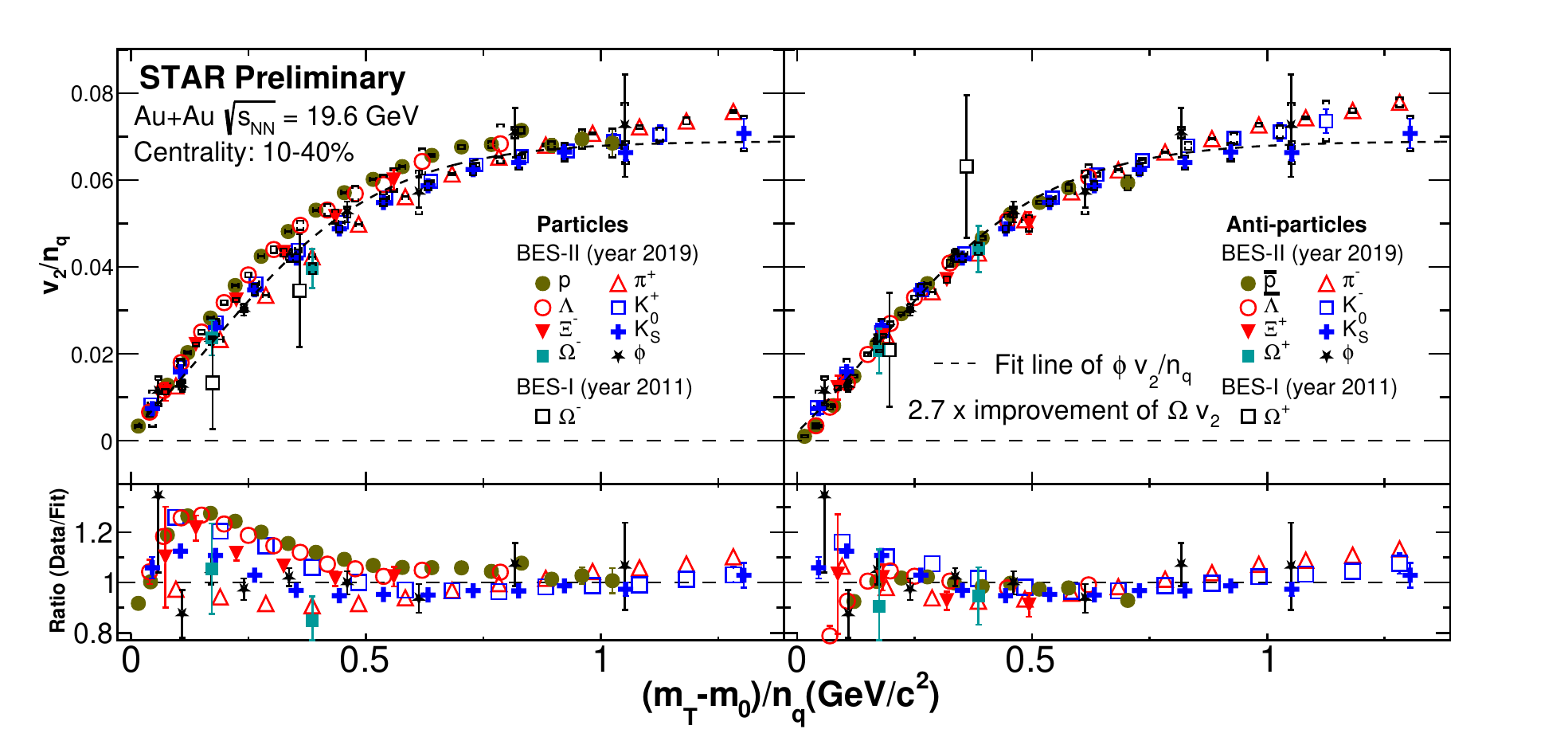}
  \caption{The number-of-constituent quark (NCQ) scaled elliptic flow, $v_{2}/n_q$ versus $(m_{T}-m_{0})/n_q$, 
  for 10-40\% central Au+Au collisions for identified particles (left pad) and corresponding anti-particles (right pad) 
  at $\sqrt{ s_{\mathrm{ NN }} }$ = 19.6 GeV where dash lines show the fit of $\phi$ mesons $v_2$.}
  \label{Fig:NCQv21040}
\end{figure}

Figure \ref{Fig:NCQv21040} shows the test of NCQ scaling of $v_2$ in centrality 10-40\% at $\sqrt{ s_{\mathrm{ NN }} }$ = 19.6 GeV.
The NCQ scaling of $v_2$ holds within 10\% for anti-particles, and 20\% for particles. A similar scaling behavior is observed for $v_3$ and holds within 15\% for anti-particles, and 30\% for particles.

\begin{figure}[htbp]
  \centering
  \includegraphics[width=3.8in,keepaspectratio]{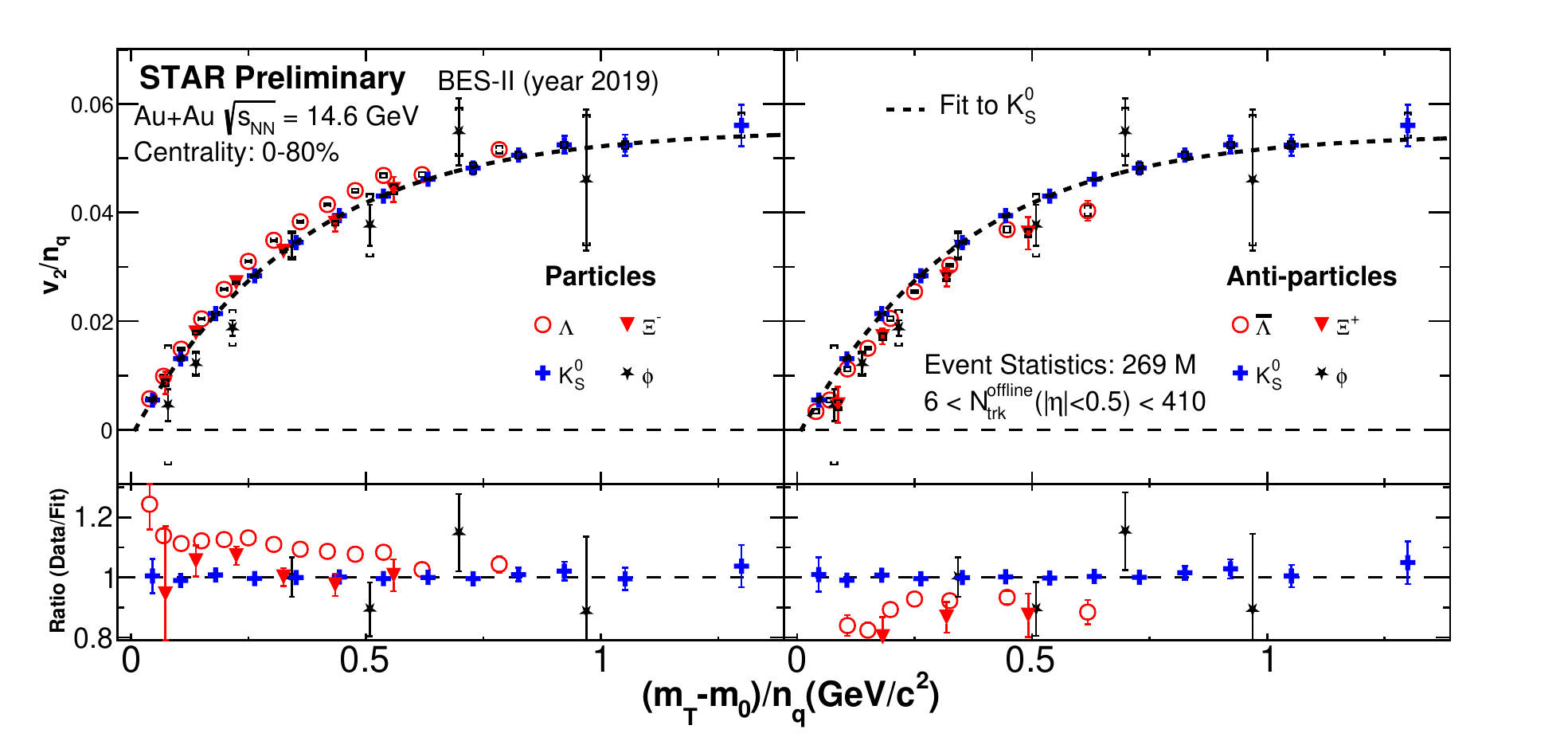}
  \caption{The number-of-constituent quark scaled elliptic flow, $v_{2}/n_q$ versus $(m_{T}-m_{0})/n_q$, 
  in 0-80\% Au+Au collisions for strange particles at $\sqrt{ s_{\mathrm{ NN }} }$ = 14.6 GeV.}
  \label{Fig:NCQv2080}
\end{figure}

The NCQ scaling reflects that the quarks are the most effective degrees of freedom in determining hadron flow 
at intermediate $p_T$, which means that the collective flow has been built up in the partonic stage 
at this collision energy. Meanwhile, 
the NCQ scaling of anti-particles works better than that of particles both for $v_2$ and $v_3$,
which may be caused by the different contributions from the produced and transported quarks.
The strange hardons and $\phi$ meson $v_2$ are measured in Au+Au collisions at BES-II 14.6 GeV, the NCQ scaling of $v_2$ holds at 20\% level shown in Fig. \ref{Fig:NCQv2080} and measurements in finer centrality bins are underway. 

\section{Summary}
\label{summary}
In summary, we report the elliptic and triangular flow of strange hadrons $K^\pm$, $K_S^0$, $\Lambda$, $\bar{\Lambda}$, $\Xi^{-}$, $\bar{\Xi}^{+}$, $\Omega^{-}$, $\bar{\Omega}^{+}$, and $\phi$ mesons at 14.6 and 19.6 GeV from BES-II. 
At $\sqrt{ s_{\mathrm{ NN }} }$ = 19.6 GeV, the NCQ scaling of $v_2$ ($v_3$) holds within 10 (15)\% for anti-particles, and within 20 (30)\% for particles.  
It indicates that the partonic collectivity could build up at this energy. Meanwhile, the NCQ scaling of anti-particles 
holds better than particles, which indicates the contribution of transported quarks in particles.
At $\sqrt{ s_{\mathrm{ NN }} }$ = 14.6 GeV, the NCQ scaling of $v_2$ holds at 20\% level in 0-80\% central Au+Au collisions.
The data taking for BES-II program (3-19.6 GeV) has completed during the year 2019 to 2021 with high statistics and detector upgrades. Such datasets will help scan the QCD phase diagram over a wide range of baryon chemical potential. 
It is expected to provide more precise differential measurements of $v_2$ and $v_3$  especially for less abundant particles: multi-strange hadrons and $\phi$ mesons. 
It will offer additional information to constrain the 
equation of state (EoS) and phase boundary of the produced QCD matter in the high baryon density region.

~\\
\begin{acknowledgement}
\textbf{Acknowledgement }
This work was supported by the National Natural Science Foundation of China [Nos. 12175084, 11890710 (11890711)], the National Key Research and Development Program of China (No. 2020YFE0202002) and the Fundamental Research Funds for the Central Universities (No. CCNU220N003).
\end{acknowledgement}

%
\bibliographystyle{woc.bst}
\bibliography{reference.bib}
%

\end{document}